**Title:** Venue-based HIV testing at sex work hotspots to reach adolescent girls and young women living with HIV: a cross-sectional study in Mombasa, Kenya

Running head: HIV testing at sex work hotspots

**Authors:** Huiting Ma[1], Linwei Wang[1], Peter Gichangi[2,4], Vernon Mochache[3], Griffins Manguro[4], Helgar K Musyoki[5], Parinita Bhattacharjee[6,7], François Cholette[8,9], Paul Sandstrom[8,9], Marissa L Becker[7], Sharmistha Mishra[1,10,11,12] on behalf of the *Transitions* Study Team

**Affiliations:**

[1]Li Ka Shing Knowledge Institute, St. Michael's Hospital, Unity Health Toronto, Toronto, Canada
[2]University of Nairobi, Nairobi, Kenya
[3]University of Maryland, Centre for International Health, Education and Biosecurity, College Park, United States
[4]International Centre for Reproductive Health-Kenya, Mombasa, Kenya
[5]National AIDS & STI Control Programme, Nairobi, Kenya
[6]Key Populations Technical Support Unit, Partners for Health and Development in Africa, Nairobi, Kenya
[7]Centre for Global Public Health, University of Manitoba, Winnipeg, Canada
[8]National HIV and Retrovirology Laboratory, JC Wilt Infectious Diseases Research Centre, Public Health Agency of Canada, Winnipeg, Canada
[9]Department of Medical Microbiology and Infectious Diseases, University of Manitoba, Winnipeg, Canada
[10]Department of Medicine, University of Toronto, Toronto, Canada
[11]Institute of Medical Science, University of Toronto, Toronto, Canada
[12]Institute of Health Policy, Management, and Evaluation, University of Toronto, Toronto, Canada

**Corresponding Author:**
Sharmistha Mishra, MD, MSc, PhD
MAP-Centre for Urban Health Solutions
St. Michael's Hospital, University of Toronto
209 Victoria St, Toronto, ON M5B 1T8
E: sharmistha.mishra@utoronto.ca
T: 416-864-5746
Fax: 416-864-5558

**Word Count:** 3506/3500

Last updated: Dec 9, 2019

**JAIDS:** 3500 words + 5 figures/tables - if more then use Appendix

**Funding**: The study was supported by the Canadian Institutes of Health Research operating grant MOP-13044 and foundation grant FDN 13455.

The authors have no funding or conflicts of interest to disclose.

# INTRODUCTION

Adolescent girls and young women (AGYW) ages 15 to 24 years face a disproportionate risk of HIV acquisition in Sub-Saharan Africa (SSA).[1] In Kenya, AGYW comprise 18.4% of the adult population but acquired 23.7% of new infections in 2017, such that by 2018 an estimated 2.6% of AGYW in Kenya were living with HIV;[2-5] yet most infections remain undiagnosed.[4] The most recent data available on AGYW suggest that in 2012, only 25% of AGYW living with HIV were diagnosed and aware of their HIV status.[4] The consequence of undiagnosed HIV among AGYW is untreated HIV, thus limiting the individual health and the population-level transmission benefits of effective antiretroviral treatment (ART).[6-8]

HIV testing serves as an entry point for HIV care with a growing recognition that differentiated strategies[9,10] – i.e. services tailored to subgroups within a population – are needed to address subgroup-specific barriers to traditional, clinic-based testing.[10,11] For example, service-related barriers reported by adolescents in SSA include stigma from healthcare providers and logistical challenges, such as costs and time for transportation to and from clinics whose hours of operation often conflict with school or employment.[11,12] Although data on differentiated strategies to improve HIV testing among AGYW remain limited,[9,10] emerging evidence suggests that venue-based testing, under the umbrella of community-based approaches, may be an effective strategy to increase HIV testing among subgroups at high risk of HIV.[13-15]

Venues refer to places where a particular subgroup may uniquely come together and socialize (schools, shopping malls, parks) and/or where people meet new sex partners.[16] For example, Herce et al. found that venue-based testing and counseling conducted as part of a survey of female sex workers, led to the new diagnosis of 63% of those living with HIV but who were previously unaware of their HIV status.[13] Venues associated with formal sex work, or sex work hotspots, are also places where AGYW, including young women who sell sex (YSW) congregate, socialize, and meet sex partners. For example, in Mombasa, Kenya, 95% of hotspots comprise venues where AGYW not engaged in sex work socialize and meet sex partners.[16] Some of these young women engage in other forms of transactional sex or casual sex and experience high prevalence of HIV-associated vulnerabilities at first sex, similar to the prevalence reported by YSW.[17]

The age of consent for HIV testing in Kenya is 15 years.[18] As in most of SSA, existing HIV testing programmes in Kenya are designed for the wider population of AGYW and/or they are designed to reach formal sex workers; they are not specifically designed to reach high risk AGYW such as those who socialize in hotspots.[18,19] Similarly, most population-based studies on HIV testing in SSA are often conducted separately for AGYW (usually via household surveys) and for female sex workers (usually restricted to those over age 18 years).[20-23] Thus, there are limited data on HIV testing patterns and undiagnosed HIV among high-risk AGYW and YSW who socialize in the same spaces. Yet studies conducted separately in each population (AGYW, female sex workers) suggest determinants of HIV testing may be similar.[20-22,24-26]

Limited data suggest YSW face similar service-related barriers to testing programmes designed for adults as those reported by the wider population of AGYW; barriers are compounded by stigma and logistical challenges related to sex work and which may also undermine access to programmes designed for AGYW in general, such as school-based testing.[11,22,27,28] Meanwhile, YSW are often excluded from sex worker programmes which provide or facilitate clinic-based HIV testing, but are designed to serve women ages 18 and over who self-identify as sex workers.[23,29] Currently in Kenya, HIV testing does not include venue-based testing at hotspots;[18] and before 2018, sex worker programmes were not allowed to provide services for women under age 18.[19,30] The consequence of vertical programmes and independently studied populations is that we do not yet know the potential value of venue-based testing at hotspots for AGYW, and whether determinants of HIV testing differ between YSW and other AGYW who frequent the same hotspots.

Among AGYW who frequent hotspots in Mombasa, Kenya, we sought to: i) compare the early elements of HIV cascade of care (diagnosis and treatment) by engagement in sex work; ii) estimate the number of AGYW living with HIV that could be newly diagnosed via hotspot-based testing; iii) compare patterns of HIV testing among AGYW by engagement in sex work; and iv) identify determinants of recent HIV testing among AGYW who frequent hotspots.

## METHODS

**Study setting and population**

We used data from hotspot enumeration and the *Transitions* Study, cross-sectional bio-behavioural survey of AGYW recruited at hotspots in Mombasa, Kenya from April to November 2015.[16,17] Survey eligibility included: cis-gender female aged 14 to 24 years who reported engaging in vaginal or anal sex at least once in their lifetime.

**Data collection**

We conducted mapping and enumeration of hotspots before survey implementation to estimate the number of AGYW aged 14 to 24 years congregating at hotspots and to generate the sampling frame as detailed in Cheuk et al.[16] We used probability proportional to estimated size of the AGYW population for sampling, and thus generated a self-weighted sample.[16,31] Within each sampled hotspot, outreach workers or a peer-educator invited potential participants, and trained interviewers screened for eligibility and administered a face-to-face structured questionnaire in English or Kiswahili. Participants were offered rapid, on-site HIV testing and counselling which was administered as per Kenya national guidelines (Appendix 1).[18] Participants who tested HIV-positive on the confirmatory rapid test were referred for HIV treatment and care. Dried blood specimens (DBS) were also collected and transferred to the National HIV and Retrovirology Laboratories in Winnipeg, Canada, which performed the serological testing with the Avioq HIV-1 Microelisa System (Avioq Inc., Research Triangle Park, NC). Participants provided written informed consent with the option to consent or decline to participate in any component of the study.[17] Data collection procedures are detailed in Becker et al.[17]

**Measures**

We classified participants as YSW if they self-identified as a sex worker or reported ever soliciting and receiving money, gifts, or other goods in exchange for sex, such that the price or commodity was negotiated prior to sex; and as women not engaged in sex work (NSW) otherwise. We used the DBS serology results to identify persons living with HIV. Participants without a DBS were excluded from our analyses of HIV cascade of care.

We defined the early stages of the HIV cascade among those living with HIV as follows: i) HIV diagnosed and aware if participants self-reported as 'HIV-positive' (those who self-reported negative or not willing to disclose or never tested for HIV were classified as undiagnosed); ii) linkage to HIV care (self-reported registration with an HIV treatment centre); and iii) currently on ART (self-reported they were currently taking antiretroviral medication).

We defined recent and lifetime HIV testing based on self-reported HIV testing with receipt of result in the year prior to the survey and ever, respectively.

We defined covariates (Appendix 2) to identify determinants of HIV testing as informed by prior literature[20,20-22,24,25,27,18] with a focus on socio-demographic, health-system engagement, sexual behaviour and risk perception; and based on data availability.

**Statistical analyses**

First, we compared the HIV cascade in YSW and NSW living with HIV.

Second, we conducted a triangulation exercise to estimate the potential number of AGYW living with HIV in Mombasa that could be newly diagnosed via hotspot-based testing if we assumed 100% test acceptance and accuracy. We used the estimated population size of AGYW who frequent hotspots in Mombasa from the 2014 mapping and enumeration;[32] and estimates of HIV prevalence and undiagnosed fraction from the current study. To estimate the feasible number of AGYW that could be newly diagnosed, we applied plausibility constraints: acceptance of rapid testing by participants who did not self-report HIV positive (measured as the proportion of participants who agreed and received rapid test when the test was offered) and the sensitivity of the rapid test against DBS results (as measured among those who received both rapid and DBS tests). We reported the potential and feasible estimates for the overall AGYW population in Mombasa who frequent at hotspots, and separately for YSW and NSW.

Third, we compared the proportion recently tested and patterns of HIV testing among YSW versus NSW. Analyses of recent HIV testing in the past year excluded participants who self-reported as 'HIV-positive' and were diagnosed with HIV more than one year before the survey.

We compared categorical variables using the chi-square tests or fisher's exact tests as appropriate, and compared continuous variables using Kruskal-Wallis tests.

To identify determinants of HIV testing, we first explored the relationship between recent testing and covariates (Appendix 2) using bivariate logistic regression among YSW and among NSW separately. To identify determinants of recent testing among AGYW who could be potentially reached by hotspot-based testing, irrespective of engagement in sex work, we performed bivariate and multivariable logistic regression on the full sample of participants. We reported the crude odds ratio (COR) and adjusted odds ratio (AOR) with 95% confidence interval (95% CI), and restricted tests of differences to variables ≥10 respondents in each cell of a predictor-outcome table. Finally, to examine the robustness of the results regarding determinants of HIV testing, we repeated our regression analyses using lifetime history of HIV testing as the outcome.

All statistical analyses and figures were executed using R version 3.4.2.

**Ethics**

The study received ethics approval from the Human Research Ethics Board at the University of Manitoba, Canada (HS16557); the Kenyatta National Hospital-University of Nairobi Ethical Review Committee, Kenya (P497/10/2017); and a research permit from the National Commission for Science, Technology and Innovation, Kenya.

# RESULTS

**Undiagnosed HIV and the HIV cascade (Figure 1)**

Of the 1,299 participants who consented to the interview (Appendix 3-**Table 1A**), 1,193 (91.8%) had DBS samples available. Participants without DBS tests were more likely to be YSW (p=0.038) and currently receiving formal education (p=0.008), but were otherwise similar to those with DBS tests (Appendix 3-**Table 1B**). Of those with a DBS test (N = 1,193), 67 (5.6%) tested HIV-positive overall. The HIV prevalence was 10.1% (37/365) among YSW and 3.6% (30/828) among NSW (p<0.001).

**Figure 1** depicts the HIV cascade. Of the 67 AGYW living with HIV, 28% (N=19) disclosed that they were diagnosed with HIV; the proportion of diagnosed and aware was 27.0% (10/37) and 30.0% (9/30) for YSW and NSW, respectively (p=0.79). Among those who were diagnosed, the majority of YSW (8/10; 80.0%) and NSW (7/9; 77.8%) self-reported to be currently on HIV treatment. A total of 13% (N=9; YSW: N=7; NSW: N=2) of AGYW living with HIV declined to tell the interviewer their HIV status, all of whom reported an HIV test in the past year. If participants who refused to report their HIV status are assumed to be diagnosed and aware, then the proportion of diagnosed and aware would represent 46.0% (17/37) and 37.0% (11/30) of YSW and NSW, respectively, living with HIV (p=0.44).

**Acceptance and sensitivity of rapid test**

A total of 1156 participants accepted rapid testing, of whom 1,124 also submitted a DBS. Using the DBS results as the gold standard, the sensitivity and specificity of the rapid test algorithm were 80.4% (95% CI): 66.9-90.2) and 99.9% (95% CI: 99.5-100.0), respectively. Among those who self-reported to be HIV-negative/not willing to disclose/never tested for HIV, 89.3% (95% CI: 87.5-91.0) accepted to have rapid testing conducted.

**Number of AGYW living with HIV who could be diagnosed via hotspot-based programs (Figure 2)**

The estimated number of AGYW frequenting hotspots in Mombasa was 15,635 (range: 12,172-19,097), of whom an estimated 6,127 (range: 4,793-7,462) were YSW.[32] Thus, using the overall HIV prevalence [5.6% (95% CI: 4.3-6.9)] and undiagnosed HIV fraction [71.6% (95% CI: 59.3-82.0)] estimates of AGYW in our study, there are an estimated 876 (range: 523-1,318) AGYW living with HIV who frequent hotspots in Mombasa, among whom an estimated 627 (range: 310-1,081) were undiagnosed. Therefore, the potential number of AGYW who could be newly diagnosed was 627 (range: 310-1,081), and the feasible number (with 89.3% test acceptance and 80.4% sensitivity) who could be newly diagnosed was 450 (range: 223-776). Thus, hotspot-based testing could feasibly reduce the undiagnosed fraction among AGYW in hotspots from 71.6% (95% CI: 59.3-82.0) to 20.2% (95% CI: 17.6-23.0).

When we stratified our triangulation by engagement in sex work, the potential and feasible numbers who could be newly diagnosed were 452 (range: 193-881) and 313 (ranges: 134-610) respectively, among YSW (Appendix 3-**Figure 2A**). Among NSW, the potential and feasible numbers who could be newly diagnosed were 240 (ranges: 93-506) and 175 (ranges 68-369), respectively (Appendix 3-**Figure 3A**).

**Profile of AGYW and patterns of HIV testing in the past year (Table 1)**

After excluding 10 participants diagnosed with HIV > 1 year prior to the survey, 1,289 were included into our analysis on patterns of recent HIV testing (**Table 1**). The median age was 19 years [inter-quartile range (IQR) 17-21]. Of the included participants, 81.0% were not aware of HIV services (74.0 % YSW vs. 84.2% NSW, p<0.001), and less than 1 in 10 (9.3%) AGYW were contacted by or registered with a non-governmental or community-based organization that provides HIV prevention services. Among those with a prior HIV test, nearly all (92.6 % YSW vs. 85.8% NSW, p=0.009) said their last test was at a public or government facility.

71.7% of participants reported a HIV test in the past year: 85.4% of YSW and 65.4% of NSW (p<0.001). HIV testing frequency in the past year was also higher among YSW than NSW (Appendix 3-**Figure 1A**). Among YSW and NSW who received an HIV test in the past year, 42.3% (146/345) and 26.6% (154/579) reported having at least two tests in the past year (p<0.001), respectively.

**Determinants of recent HIV testing among AGYW who frequent hotspots (Tables 2 and 3)**

Determinants of recent HIV testing were similar among YSW and NSW (**Table 2**).

**Table 3** provides the determinants of recent HIV testing among AGYW who frequent hotspots. The size and direction of determinants identified in bivariate analysis persisted after adjusting for engagement in sex work and other covariates. Older age [AOR (95% CI): 1.5 (1.2-2.1)], higher education attainment [AOR (95% CI): 1.4 (1.0-2.0)], and longer duration of sexual activity [AOR (95% CI): 1.4 (1.0-1.8) were independently associated with receiving an HIV test in the past one year (**Table 3**). Prior engagement with the healthcare system due to a history of pregnancy or treatment for a sexually transmitted infection in the past year, were also independently associated with HIV testing [AOR (95% CI): 1.8 (1.3-2.5), AOR (95% CI): 1.9 (1.3-2.9), respectively]; and awareness of sex worker programmes [AOR (95% CI): 1.7 (1.2-2.5)]. In contrast, participants who were in formal education at the time of the survey (vs. being out of school) were less likely to have been tested in the past year [AOR (95% CI): 0.7 (0.5-1.0)] (**Table 3**). After adjusting for these determinants of recent HIV testing, YSW were two-fold more likely to have tested for HIV in the past year [AOR (95% CI): 2.1 (1.6-3.1)]. Self-assessed HIV risk was associated with a greater likelihood of HIV testing on univariate analyses, but not after adjusting for other determinants (**Table 3**). Sensitivity analyses identified similar determinants for lifetime HIV testing (Appendix 3-**Table 2A and Table 3A**).

## DISCUSSION

We identified a large unmet need in HIV diagnoses among AGYW who frequent hotspots in Mombasa, Kenya. Although 86% of AGYW reported a lifetime history of HIV testing, only 72% were tested in the previous year, and less than 1 in 3 AGYW living with HIV were diagnosed and aware of their status. YSW were more likely to be living with HIV, and were three-fold more likely to test for HIV in the past year, and would do so more frequently, than AGYW who did not sell sex. However, the prevalence of undiagnosed HIV, and the determinants of HIV testing were similar across AGYW irrespective of whether or not they were engaged in sex work. Applying a hotspot-based strategy of onsite HIV testing with existing rapid tests could realistically and newly diagnose 51.4% of AGYW living with HIV who socialize at hotspots.

Our findings suggest that hotspots comprise subsets of AGYW with disproportionately high risk of HIV and poor access and/or uptake of HIV testing services; and similar to findings of disproportionate risks among AGYW who socialize at other types of venues (bars, hotels, transportation hubs) in East Africa.[33] As shown with other populations within Kenya, once diagnosed with HIV, the proportion who go on to receive ART is high,[34] suggesting that diagnosis remains a critical gap in the HIV cascade. Extrapolation of Mombasa-specific estimates of undiagnosed HIV among AGYW at hotspots to the national-level (via multiplying by the relative difference in AGYW population size in Kenya as a whole [4,066,888] versus Mombasa [134,885]) suggests that in Kenya, there could be an estimated 15,789 to 39,729 AGYW living with HIV who frequent hotspots and would benefit from ART; and of whom nearly 9,347 and 32,593 currently remain undiagnosed and unaware.[32,35]

The discrepancy between the relatively high proportion of participants recently tested for HIV yet low proportion diagnosed may reflect inadequate frequency and timing of tests in relation to changes in HIV risk over time or age. Local programmes in Kenya offer HIV testing every three months for sex workers, and annual testing for AGYW in general.[18,36] In our study, only 9.3% of YSW who tested in the past year did so at least 4 times; thus most YSW tested less frequently than what is recommended for women engaged in sex work.[18,19] The optimal frequency and timing of tests may also need to be adapted to the changing experiences and exposures in an AGYW's sexual life-course, and should be facilitated by approaches that enhance an individual's agency over testing – such as HIV self-testing.[9] In our study, 10% of YSW were already living with HIV and yet had only been in sex work for a median of 2 years[17] – suggesting either a high prevalence of HIV prior to entering sex work and/or high incidence of HIV within the first two years of sex work. The latter in particular means that testing frequency may need to be even higher during the early period of sex work.

High levels of recent HIV testing and high undiagnosed fraction could also result from the sensitivity (81%) of the rapid tests used in the Kenya national standard protocols. If we apply the false-negative rate of the rapid test to AGYW living with HIV tested in the last year, the undiagnosed fraction is still high at 62.7%. The discrepancy between recent testing and undiagnosed fraction is also important in the context of evaluating HIV testing strategies, many of which use test uptake as the main outcome.[25,37] Thus, our findings suggest that monitoring and evaluation of testing strategies should also measure undiagnosed fraction at the population-level rather than just the proportion tested in the previous year.

Compared with NSW, YSW were more likely to have tested for HIV in the past year; findings which correspond with higher rates of HIV testing among Kenyan women engaged in sex work compared to the wider population.[4,38] However, YSW and NSW recruited from hotspots shared several determinants of HIV testing. Thus, if a hotspot-based testing strategy in Mombasa was to also deploy risk-profiling to prioritize those least likely to have tested recently, it could use the same profiles for YSW and for NSW.

To date, venue-based strategies deployed for AGYW have been restricted to mobile-outreach at parks and entertainment venues; all of which suggest increased uptake of HIV testing among adolescents.[14,15,39-41] Our findings suggest hotspot-based testing strategies, such as that deployed as part of the *Transitions* Study, represent an untapped opportunity to increase HIV diagnoses among AGYW living with HIV. Indeed, a population-based strategy to deliver testing services to hotspots may not require individuals to self-identify as engaging in sex work; and thus provide an avenue to converge outreach and service delivery from the disparate pillars of adolescent and sex work programmes. Recommendations for testing – across key populations including AGYW – include the provision of a "safe space", testing free of coercion and employing approaches that

address stigma and discrimination related to sex work in general, and to sexual activity among youth.[42]

Study limitations include the use of self-reported data collected via face-to-face interviews which may be prone to measurement and social desirability bias, respectively. Estimates of the cascade of HIV care are also limited by the 16% of participants without reference DBS tests, and the 13% of AGYW living with HIV who did not wish to disclose their status to the interviewer. However, limitations on restricting our study population to those with DBS may be mitigated by the similar profile of participants with and without DBS.

In conclusion, there remains a large unmet need in the early elements of the HIV cascade among a particularly high-risk subset of AGYW in Kenya. Reaching AGYW via hotspot-based HIV testing strategies may reach higher risk AGYW and fill gaps left by traditional HIV prevention and testing services.

# ACKNOWLEDGEMENTS

The authors thank all the women who participated in this study and acknowledge the efforts of the *Transitions* Study team and their partners. We thank the International Centre for Reproductive Health, National AIDS Control Council and the National AIDS & STI Control Programme, Institute for Economics and Forecasting National Academy of Sciences of Ukraine and Ukrainian Institute for Social Research after Oleksandr Yaremenko. We thank Japheth Kioko and Shem Kaosa from Kenya Technical Support Unit for data-entry and data-cleaning support. We thank Dr. Eve Cheuk (University of Manitoba) who led overall coordination of data collection, and Dr. Shajy Isac (Karnataka Health Promotion Trust) who led the mapping and enumeration. We thank Kristy Yiu (Unity Health Toronto) for proof-reading and helping with the submission.

SM is supported by a Canadian Institutes of Health Research (CIHR) and Ontario HIV Treatment Network New Investigator Award. MB is supported by a CIHR New Investigator Award.

# AUTHOR CONTRIBUTIONS

HMa, LW, and SM conceptualized and designed the study and developed the plan of analyses. MB, SM, HMu, PB developed the study tools; MB, PG, GM, HMu, and PB led the hotspot enumeration and *Transitions* study data collection. FC and PS led the serological testing and developed the reference testing algorithms. All authors contributed to interpretation of results and manuscript editing. SM and HMa drafted the manuscript. HMa conducted the analyses with input from LW and SM.

**Role of funding source**

# TABLES

**Table 1. Characteristics of the study participants aged 14-24 years by engagement in sex work in Mombasa, Kenya (N = 1289).**

| Characteristics (N (%)) | Overall[e] (N = 1289) | YSW (N = 404) | NSW (N = 885) | p-value |
|---|---|---|---|---|
| **Socio-demographic characteristics** | | | | |
| **Type of recruitment hotspot** | | | | |
| Physical establishments[a] | 1060 (82.2%) | 344 (85.1%) | 716 (80.9%) | 0.07 |
| Public spaces[b] | 229 (17.8%) | 60 (14.9%) | 169 (19.1%) | |
| **Age in years** | | | | |
| 14-18 | 521 (40.4%) | 117 (29.0%) | 404 (45.6%) | < 0.001 |
| 19-24 | 768 (59.6%) | 287 (71.0%) | 481 (54.4%) | |
| **The highest education level** | | | | |
| Did not complete primary school | 300 (23.3%) | 121 (30.0%) | 179 (20.2%) | < 0.001 |
| Completed primary school | 666 (51.7%) | 209 (51.7%) | 457 (51.6%) | |
| Completed secondary school or higher | 323 (25.1%) | 74 (18.3%) | 249 (28.1%) | |
| **Currently receiving formal education** | 264 (20.5%) | 33 (8.2%) | 231 (26.1%) | < 0.001 |
| **Health-system engagement** | | | | |
| **Ever pregnant** | 485 (37.6%) | 230 (56.9%) | 255 (28.8%) | < 0.001 |
| **Treated STI last 1 year** | 219 (17.0%) | 89 (22.0%) | 130 (14.7%) | 0.001 |
| **Programme engagement** | | | | |
| Not aware of HIV services | 1044 (81.0%) | 299 (74.0%) | 745 (84.2%) | < 0.001 |
| Awareness of HIV services | 126 (9.8%) | 47 (11.6%) | 79 (8.9%) | |
| Ever contacted by peers/staff from an NGO/CBO | 55 (4.3%) | 21 (5.2%) | 34 (3.8%) | |
| Registered with NGO/CBO | 64 (5.0%) | 37 (9.2%) | 27 (3.1%) | |
| **Ever received an HIV test** | 1111 (86.2%) | 379 (93.8%) | 732 (82.7%) | < 0.001 |
| **Tested for HIV in the last 1 year** | 924 (71.7%) | 345 (85.4%) | 579 (65.4%) | < 0.001 |
| **Last HIV testing location** | | | | |
| Public/government facility | 979 (88.1%) | 351 (92.6%) | 628 (85.8%) | 0.009 |
| NGO/CBO through outreach | 41 (3.7%) | 10 (2.6%) | 31 (4.2%) | |
| Private facility | 22 (2.0%) | 4 (1.1%) | 18 (2.5%) | |
| Other/do not recall | 69 (6.2%) | 14 (3.7%) | 55 (7.5%) | |
| **Sexual behavior and risk perception** | | | | |
| **Duration of sexual activity[c]** | | | | |
| <2 years | 432 (33.5%) | 63 (15.6%) | 369 (41.7%) | < 0.001 |
| >=2 years | 857 (66.5%) | 341 (84.4%) | 516 (58.3%) | |
| **Duration in sex work** | | | | |
| <2 years | 199 (49.3%) | 199 (49.3%) | | -- |
| >=2 years | 205 (50.7%) | 205 (50.7%) | | |
| **Self-assessed risk of HIV acquisition[d] (N=1282)** | | | | |
| No risk at all/small/unsure | 745 (58.1%) | 181 (45.0%) | 564 (64.1%) | < 0.001 |
| Moderate/great | 537 (41.9%) | 221 (55.0%) | 316 (35.9%) | |

Abbreviations: CBO (community-based organization); NGO (non-governmental organization); NSW (young women not engaged in sex work); STI (sexually transmitted infection); YSW (young women who sell sex)
[a]Physical establishments hotspots include bars, night clubs, hotels, guest houses, lodges, restaurants, local brew dens, sex dens and brothels
[b]Public spaces hotspots include streets and other public places
[c]N=55/1289 missing was imputed by adjusting for age at the interview
[d]Excluding individuals who disclosed they are living with HIV
[e]Excluding individuals who were diagnosed with HIV >1 year ago

**Table 2. Factors associated with HIV testing in the past year among adolescent girls and young women aged 14-24 years by engagement in sex work in Mombasa, Kenya (N=1289).**

| Characteristics | YSW (N=404) Yes (%) | Crude OR (95% CI) | p-value | NSW (N=885) Yes (%) | Crude OR (95% CI) | p-value |
|---|---|---|---|---|---|---|
| **Socio-demographic characteristics** | | | | | | |
| **Type of recruitment hotspot** | | | | | | |
| Physical establishments[a] | 49 (81.7%) | 1.4 (0.6 - 2.8) | 0.38 | 117 (69.2%) | 0.8 (0.6 - 1.2) | 0.25 |
| Public spaces[b] | 296 (86.0%) | Ref | | 462 (64.5%) | Ref | |
| **Age in years** | | | | | | |
| 14-18 | 88 (75.2%) | Ref | | 224 (55.4%) | Ref | |
| 19-24 | 257 (89.5%) | 2.8 (1.6 - 5.0) | < 0.001 | 355 (73.8%) | 2.3 (1.7 - 3.0) | < 0.001 |
| **The highest education level** | | | | | | |
| Did not complete primary school | 92 (76%) | Ref | | 108 (60.3%) | Ref | |
| Completed primary school | 180 (86.1%) | 2.0 (1.1 - 3.5) | 0.022 | 291 (63.7%) | 1.2 (0.8 - 1.6) | 0.43 |
| Completed secondary school or higher | 73 (98.6%) | 23.0 (4.7 - 414.8) | 0.002 | 180 (72.3%) | 1.7 (1.1 - 2.6) | 0.010 |
| **Currently receiving formal education** | | | | | | |
| No | 315 (84.9%) | Ref | | 461 (70.5%) | Ref | |
| Yes | 30 (90.9%) | 1.8 (0.6 - 7.6) | 0.36 | 118 (51.1%) | 0.4 (0.3 - 0.6) | < 0.001 |
| **Health-system engagement** | | | | | | |
| **Ever pregnant** | | | | | | |
| No | 141 (81%) | Ref | | 377 (59.8%) | Ref | |
| Yes | 204 (88.7%) | 1.8 (1.1 - 3.2) | 0.033 | 202 (79.2%) | 2.6 (1.8 - 3.6) | < 0.001 |
| **Treated STI last 1 year** | | | | | | |
| No | 259 (82.2%) | Ref | | 479 (63.4%) | Ref | |
| Yes | 86 (96.6%) | 6.2 (2.2 - 25.9) | 0.003 | 100 (76.9%) | 1.9 (1.3 - 3.0) | 0.003 |
| **Programme engagement** | | | | | | |
| Not aware of HIV services | 245 (81.9%) | Ref | | 479 (64.3%) | Ref | |
| Awareness of HIV services | 44 (93.6%) | 3.2 (1.1 - 13.7) | 0.057 | 52 (65.8%) | 1.1 (0.7 - 1.8) | 0.79 |
| Ever contacted by peers/staff from an NGO/CBO | 20 (95.2%) | 4.4 (0.9 - 80.0) | 0.15 | 28 (82.4%) | 2.6 (1.1 - 7.0) | 0.037 |
| Registered with NGO/CBO | 36 (97.3%) | 7.9 (1.7 - 142.5) | 0.043 | 20 (74.1%) | 1.6 (0.7 - 4.1) | 0.30 |
| **Sexual behavior and risk perception** | | | | | | |
| **Duration of sexual activity[c]** | | | | | | |
| <2 years | 48 (76.2%) | Ref | | 207 (56.1%) | Ref | |
| >=2 years | 297 (87.1%) | 2.1 (1.1 - 4.0) | 0.023 | 372 (72.1%) | 2.0 (1.5 - 2.7) | < 0.001 |
| **Duration in sex work** | | | | | | |
| <2 years | 171 (85.9%) | Ref | | -- | -- | -- |
| >=2 years | 174 (84.9%) | 0.9 (0.5 - 1.6) | 0.76 | -- | -- | -- |
| **Self-assessed risk of HIV acquisition[d]** | | | | | | |
| No risk at all/small/unsure | 155 (85.6%) | Ref | | 361 (64.0%) | Ref | |
| Moderate/great | 189 (85.5%) | 1.0 (0.6 - 1.7) | 0.97 | 214 (67.7%) | 1.2 (0.9 - 1.6) | 0.27 |

Abbreviations: CBO (community-based organization); NGO (non-governmental organization); NSW (young women not engaged in sex work); STI (sexually transmitted infection); YSW (young women who sell sex)
[a] Physical establishments hotspots include bars, night clubs, hotels, guest houses, lodges, restaurants, local brew dens, sex dens and brothels
[b] Public spaces hotspots include streets and other public places
[c] N=55/1289 missing was imputed by adjusting for age at the interview
[d] Excluding individuals who disclosed they are living with HIV
[e] Excluding individuals who were diagnosed with HIV >1 year ago

**Table 3.** Univariate and multivariable analyses of factors associated with HIV testing in the past year among adolescent girls and young women aged 14-24 years in Mombasa, Kenya (N=1289).

| Characteristics | HIV testing in the past year[g] | | | |
|---|---|---|---|---|
| | Crude OR (95% CI) | p-value | Adjusted OR (95% CI)[f] | p-value |
| **Socio-demographic characteristics** | | | | |
| **Engagement in sex work** | | | | |
| No | Ref | < 0.001 | Ref | < 0.001 |
| Yes | 3.1 (2.3 - 4.2) | | 2.1 (1.6 - 3.1) | |
| **Type of recruitment hotspot[f]** | | | | |
| Physical establishments[a] | 1.0 (0.7 - 1.3) | 0.77 | -- | |
| Public spaces[b] | Ref | | -- | |
| **Age in years** | | | | |
| 14-18 | Ref | < 0.001 | Ref | 0.003 |
| 19-24 | 2.6 (2.1 - 3.4) | | 1.5 (1.2 - 2.1) | |
| **Completed primary school** | | | | |
| No | Ref | 0.028 | Ref | 0.003 |
| Yes | 1.4 (1.0 - 1.8) | | 1.4 (1.0 - 2.0) | |
| **Currently receiving formal education** | | | | |
| No | Ref | < 0.001 | Ref | 0.044 |
| Yes | 0.4 (0.3 - 0.5) | | 0.7 (0.5 - 1.0) | |
| **Health-system engagement** | | | | |
| **Ever pregnant** | | | | |
| No | Ref | < 0.001 | Ref | < 0.001 |
| Yes | 2.8 (2.2 - 3.8) | | 1.8 (1.3 - 2.5) | |
| **Treated STI last 1 year** | | | | |
| No | Ref | < 0.001 | Ref | 0.002 |
| Yes | 2.5 (1.7 - 3.8) | | 1.9 (1.3 -2.9) | |
| **Awareness of HIV services** | | | | |
| No | Ref | < 0.001 | Ref | 0.004 |
| Yes | 2.0 (1.4 - 2.8) | | 1.7 (1.2 - 2.5) | |
| **Sexual behavior and risk perception** | | | | |
| **Duration of sexual activity[c]** | | | | |
| <2 years | Ref | < 0.001 | Ref | 0.033 |
| >=2 years | 2.5 (1.9 - 3.2) | | 1.4 (1.0 - 1.8) | |
| **Duration in sex work[g]** | | | | |
| <2 years | -- | | -- | |
| >=2 years | -- | | -- | |
| **Self-assessed risk of HIV acquisition[e]** | | | | |
| No risk at all/small/unsure | Ref | 0.024 | Ref | 0.73 |
| Moderate/great | 1.3 (1.0 - 1.7) | | 1.1 (0.8 – 1.4) | |

Abbreviations: OR (odds ratio); STI (sexually transmitted infection)
[a]Physical establishments hotspots include bars, night clubs, hotels, guest houses, lodges, restaurants, local brew dens, sex dens and brothels
[b]Public spaces hotspots include streets and other public places
[c]N=55/1289 missing was imputed by adjusting for age at the interview
[d]Excluding individuals who disclosed they are living with HIV
[e]Excluding individuals who were diagnosed with HIV >1 year ago
[f]Not included covariates in multivariable analysis if significance level > 0.1 in univariate analysis
[g]Not included in univariable and multivariable analysis due to duration in sex work only applied to YSW

# FIGURES

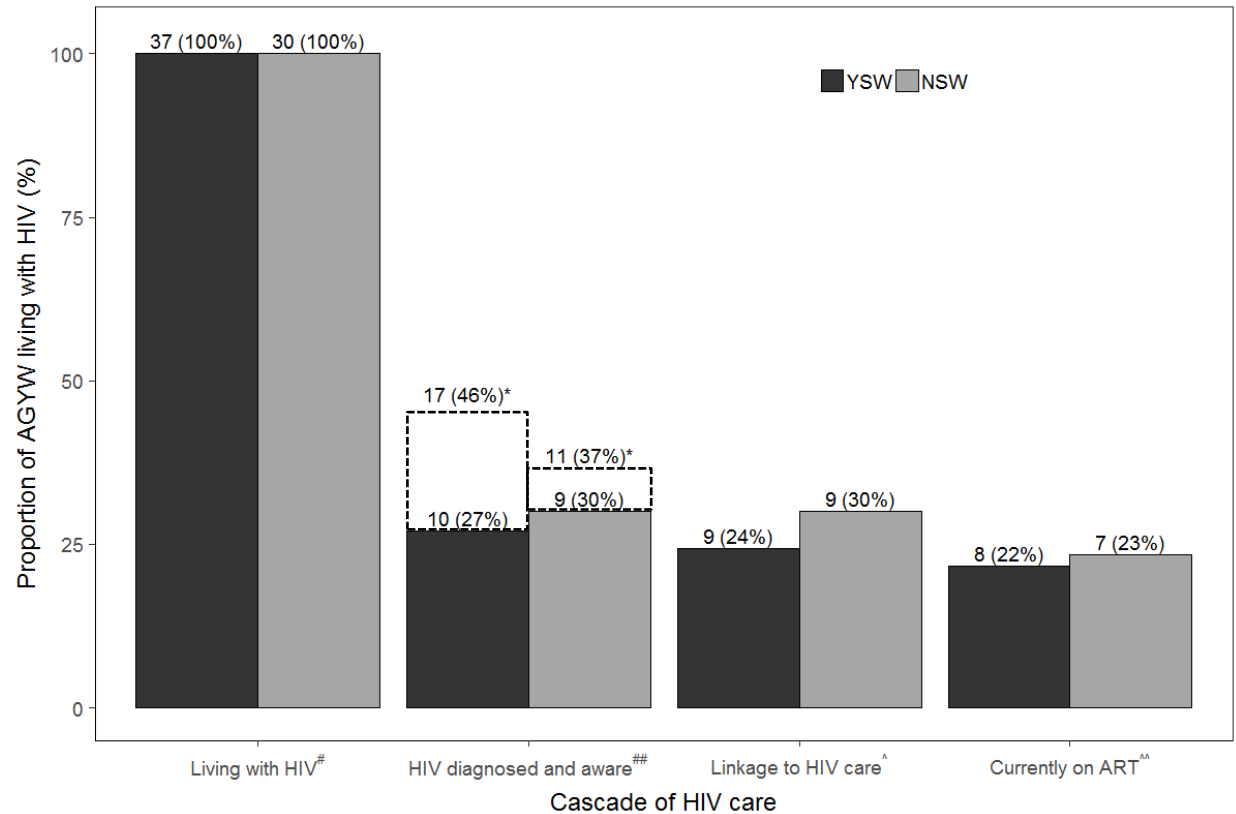

**Figure 1. Cascade of HIV care among adolescent girls and young women aged 14-24 years living with HIV by engagement in sex work in Mombasa, Kenya (N=67).**
Abbreviations: AGYW (adolescent girls and young women); ART (antiretroviral therapy); NSW (young women not engaged in sex work); YSW (young women who sell sex)
*Based on the assumption that participants who were not willing to disclose their HIV-status were living with HIV and were aware of their status
#Based on DBS serology results
##Self-reported as 'HIV-positive' (those who self-reported as HIV-negative or not willing to disclosure or never tested for HIV were classified as undiagnosed)
^Self-reported registration with an HIV treatment centre
^^Self-reported that they were currently taking antiretroviral medication

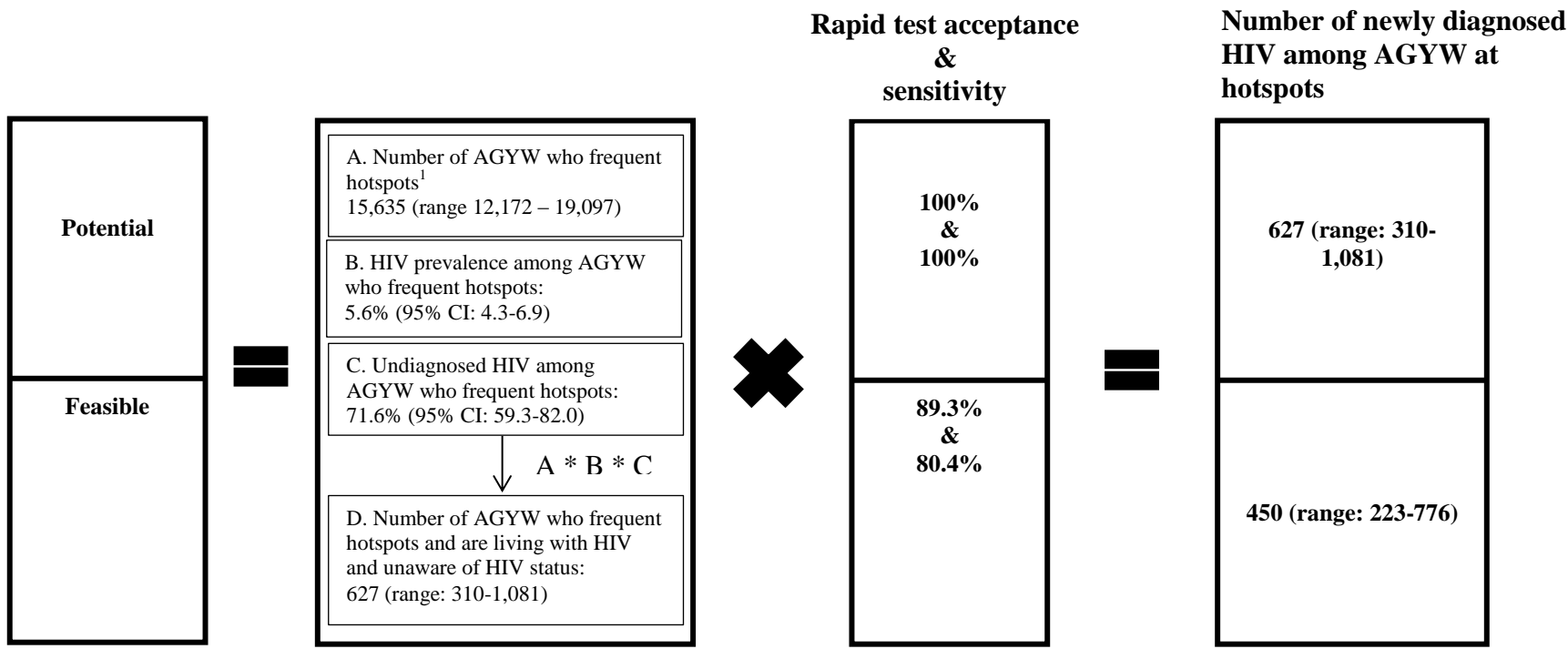

**Figure 2. Triangulating the number of adolescent girls and young women living with HIV who could be diagnosed via hotspot-based HIV testing strategy in Mombasa, Kenya.**
Abbreviation: AGYW (adolescent girls and young women).
[1]Cheuk E, Isac S, Musyoki H, et al. Informing HIV Prevention Programs for Adolescent Girls and Young Women: A Modified Approach to Programmatic Mapping and Key Population Size Estimation. *JMIR Public Health Surveill.* 2019;5(2):e11196.

# APPENDIX 1

**Algorithm for rapid HIV testing at sex work venues (hotspots)**
A trained clinical officer/nurse counsellor performed the tests along with pre- and post-test counselling, and service-referral in accordance with the Ministry of Health of Kenya.[1] The rapid tests were performed at or near the hotspot in a closed (confidential) room and at the same place as the face-to-face interview.

The first test was KHB HIV (1+2) Antibody (Colloidal Gold) Rapid Test (Shanghai Kehua Bio-engineering Co., Ltd, Shanghai, China); the second test was First Response HIV 1-2-O Rapid Whole Blood Test (Premier Medical Corporation Private Limited, Mumbai, India); and the tie-breaker test was Uni-Gold™ HIV Test (Trinity Biotech Plc, Bray, Ireland). If the results of first test were negative, the participant was informed of the negative results and was referred to the appropriate HIV prevention services. If the results of first test were positive, the second test was performed. If the results of second test were also positive, the participant was informed of the positive results and was referred to the appropriate HIV treatment services. If the results of second test were negative, the results were classified as discordant and a third blood sample was collected for the tie-breaker test. Participants were informed of their rapid test results at the same visit.

# APPENDIX 2

**Covariate Definitions**

To identify determinants of recent HIV testing among adolescent girls and young women who socialize at sex work hotspots, we focused on socio-demographic, health-system engagement, sexual behaviour, and risk perception.

For socio-demographic characteristics, we included the type of hotspot from where the participant was recruited (physical establishments [bars, night clubs, hotels, guest houses, lodges, restaurants, local brew dens, sex dens and brothels]; public spaces [streets and other public places such as beach, park etc.]; age at time of interview (14-18; 19-24 years); educational attainment (did not complete primary school; completed primary school; completed secondary school or higher) and currently receiving formal education. For health-system engagement, we included history of pregnancy; treatment for a bacterial sexually transmitted infection in the previous year; and engagement with sex worker programmes (awareness of an HIV programme, ever contacted by peers/staffs from a non-governmental organization/community-based organization, registration with HIV-prevention programme) as these questions were asked of all study participants. Within sexual behavior and risk perception, we included duration of sexual activity ($< 2$ years; $>= 2$ years); duration of sex work ($< 2$ years; $>= 2$ years) among women who sell sex (YSW) only; self-assessed risk of HIV acquisition for those who were not diagnosed with HIV (no risk at all/small/unsure; moderate/great).

# APPENDIX 3

**Table 1A. Characteristics of the study participants aged 14-24 years by engagement in sex work in Mombasa, Kenya (N = 1299).**

| Characteristics (N (%)) | Overall (N = 1299) | YSW (N = 408) | NSW (N = 891) | p-value |
|---|---|---|---|---|
| **Socio-demographic characteristics** | | | | |
| **Type of recruitment hotspot** | | | | |
| Physical establishments[a] | 1069 (82.0%) | 348 (85.3%) | 721 (80.9%) | 0.06 |
| Public spaces[b] | 230 (18.0%) | 60 (14.7%) | 170 (19.1%) | |
| **Age in years** | | | | |
| 14-18 | 522 (40.0%) | 117 (28.7%) | 405 (45.5%) | < 0.001 |
| 19-24 | 777 (60.0%) | 291 (71.3%) | 486 (54.5%) | |
| **The highest education level** | | | | |
| Did not complete primary school | 305 (23.5%) | 124 (30.4%) | 181 (20.3%) | < 0.001 |
| Completed primary school | 671 (51.7%) | 210 (51.5%) | 461 (51.7%) | |
| Completed secondary school or higher | 323 (24.9%) | 74 (18.1%) | 249 (27.9%) | |
| **Currently receiving formal education** | 266 (20.5%) | 33 (8.1%) | 233 (26.2%) | < 0.001 |
| **Health-system engagement** | | | | |
| **Ever pregnant** | 493 (38.0%) | 234 (57.4%) | 259 (29.1%) | < 0.001 |
| **Treated STI last 1 year** | 223 (17.0%) | 91 (22.3%) | 132 (14.8%) | 0.001 |
| **Programme engagement** | | | | |
| Not aware of HIV services | 1052 (81.0%) | 303 (74.3%) | 749 (84.1%) | < 0.001 |
| Awareness of HIV services | 127 (9.8%) | 47 (11.5%) | 80 (9%) | |
| Ever contacted by peers/staff from an NGO/CBO | 56 (4.3%) | 21 (5.1%) | 35 (3.9%) | |
| Registered with NGO/CBO | 64 (4.9%) | 37 (9.1%) | 27 (3%) | |
| **Ever received an HIV test** | 1121 (86.0%) | 383 (93.9%) | 738 (82.8%) | < 0.001 |
| **Tested for HIV in the last 1 year[c]** | 924 (72.0%) | 345 (85.4%) | 579 (65.4%) | < 0.001 |
| **Last HIV testing location** | | | | |
| Public/government facility | 989 (88.0%) | 355 (92.7%) | 634 (85.9%) | 0.008 |
| NGO/CBO through outreach | 41 (4.0%) | 10 (2.6%) | 31 (4.2%) | |
| Private facility | 22 (2.0%) | 4 (1.0%) | 18 (2.4%) | |
| Other/Do not recall | 69 (6.0%) | 14 (3.7%) | 55 (7.5%) | |
| **Sexual behavior and risk perception** | | | | |
| **Duration of sexual activity[d]** | | | | |
| <2 years | 434 (33.4%) | 63 (15.4%) | 371 (41.6%) | < 0.001 |
| >=2 years | 865 (66.6%) | 345 (84.6%) | 520 (58.4%) | |
| **Duration in sex work** | | | | |
| <2 years | 200 (49.0%) | 200 (49.0%) | -- | |
| >=2 years | 208 (51.0%) | 208 (51.0%) | -- | |
| **Self-assessed risk of HIV acquisition[e] (N=1283)** | | | | |
| No risk at all/small/unsure | 745 (58.0%) | 222 (55.1%) | 316 (35.9%) | < 0.001 |
| Moderate/Great | 538 (42.0%) | 181 (44.9%) | 564 (64.1%) | |

Abbreviations: CBO (community-based organization);  ; NGO (non-governmental organization); NSW (young women not engaged in sex work); STI (sexually transmitted infection); YSW (young women who sell sex)
[a]Physical establishments hotspots include bars, night clubs, hotels, guest houses, lodges, restaurants, local brew dens, sex dens and brothels
[b]Public spaces hotspots include streets and other public places
[c]Excluding individuals who were diagnosed with HIV >1 year ago
[d]N=55/1299 missing was imputed by adjusting for age at the interview
[e]Excluding individuals who disclosed they are living with HIV

**Table 1B. Characteristics of study participants age 14-24 years in Mombasa, by availability of DBS HIV test results (N =1299).**

| Characteristics (N (%)) | Overall (N = 1299) | Accepted (N = 1193) | Declined (N = 106) | p-value |
|---|---|---|---|---|
| **Socio-demographic characteristics** | | | | |
| **Engagement in sex work** | | | | |
| No | 891 (68.6%) | 828 (69.4%) | 63 (59.4%) | 0.038 |
| Yes | 408 (31.4%) | 365 (30.6%) | 43 (40.6%) | |
| **Type of recruitment hotspot** | | | | |
| Physical establishments[a] | 1069 (82.3%) | 987 (82.7%) | 82 (77.4%) | 0.18 |
| Public spaces[b] | 230 (17.7%) | 206 (17.3%) | 24 (22.6%) | |
| **Age in years** | | | | |
| 14-18 | 522 (40.2%) | 486 (40.7%) | 36 (34%) | 0.18 |
| 19-24 | 777 (59.8%) | 707 (59.3%) | 70 (66%) | |
| **The highest education level** | | | | |
| Did not complete primary school | 305 (23.5%) | 281 (23.6%) | 24 (22.6%) | 0.29 |
| Completed primary school | 671 (51.7%) | 622 (52.1%) | 49 (46.2%) | |
| Completed secondary school or higher | 323 (24.9%) | 290 (24.3%) | 33 (31.1%) | |
| **Currently receiving formal education** | 266 (20.5%) | 233 (19.5%) | 33 (31.1%) | 0.008 |
| **Health-system engagement** | | | | |
| **Ever pregnant** | 493 (38%) | 454 (38.1%) | 39 (36.8%) | 0.84 |
| **Treated STI last 1 year** | 223 (17.2%) | 199 (16.7%) | 24 (22.6%) | 0.14 |
| **Programme engagement** | | | | |
| Not aware of HIV services | 1052 (81%) | 969 (81.2%) | 83 (78.3%) | 0.74 |
| Awareness of HIV services | 127 (9.8%) | 116 (9.7%) | 11 (10.4%) | |
| Ever contacted by peers/staff from an NGO/CBO | 56 (4.3%) | 51 (4.3%) | 5 (4.7%) | |
| Registered with NGO/CBO | 64 (4.9%) | 57 (4.8%) | 7 (6.6%) | |
| **Ever received an HIV test** | 1121 (86.3%) | 1031 (86.4%) | 90 (84.9%) | 0.66 |
| **Tested for HIV in the last 1 year** | 924 (71.7%) | 849 (71.6%) | 75 (72.1%) | 1.00 |
| **Last HIV testing location** | | | | |
| Public/government facility | 989 (88.2%) | 910 (88.3%) | 79 (87.8%) | 0.96 |
| NGO/CBO through outreach | 41 (3.7%) | 38 (3.7%) | 3 (3.3%) | |
| Private facility | 22 (2%) | 20 (1.9%) | 2 (2.2%) | |
| Other/Do not recall | 69 (6.2%) | 63 (6.1%) | 6 (6.7%) | |
| **Sexual behavior and risk perception** | | | | |
| **Duration of sexual activity[c]** | | | | |
| <2 years | 434 (33.4%) | 395 (33.1%) | 39 (36.8%) | 0.45 |
| >=2 years | 865 (66.6%) | 798 (66.9%) | 67 (63.2%) | |
| **Duration in sex work (YSW only)** | | | | |
| <2 years | 200 (49%) | 181 (49.6%) | 19 (44.2%) | 0.52 |
| >=2 years | 208 (51%) | 184 (50.4%) | 24 (55.8%) | |
| **Self-assessed risk of HIV acquisition[d] (N=1283)** | | | | |
| No risk at all/small/unsure | 745 (58.1%) | 693 (58.8%) | 52 (50.0%) | 0.10 |
| Moderate/Great | 538 (41.9%) | 486 (41.2%) | 52 (50.0%) | |

Abbreviations: CBO (community-based organization); NGO (non-governmental organization); STI (sexually transmitted infection)

[a]Physical establishments hotspots include bars, night clubs, hotels, guest houses, lodges, restaurants, local brew dens, sex dens and brothels

[b]Public spaces hotspots include streets and other public places

[c]N=55/1299 missing was imputed by adjusting for age at the interview

[d]Excluding individuals who disclosed they are living with HIV

**Table 2A. Factors associated with the history of HIV testing among adolescent girls and young women aged 14-24 years by engagement in sex work in Mombasa, Kenya (N=1299).**

| | Proportion with ever received HIV test | | | | | |
|---|---|---|---|---|---|---|
| | YSW (N=408) | | | NSW (N=891) | | |
| Characteristics | Yes (%) | Crude OR (95% CI) | p-value | Yes (%) | Crude OR (95% CI) | p-value |
| **Socio-demographic characteristics** | | | | | | |
| **Type of recruitment hotspot** | | | | | | |
| Physical establishments[a] | 55 (91.7%) | 1.5 (0.5 - 3.9) | 0.44 | 149 (87.6%) | 0.6 (0.4 – 1.0) | 0.07 |
| Public spaces[b] | 328 (94.3%) | Ref | | 589 (81.7%) | Ref | |
| **Age in years** | | | | | | |
| 14-18 | 100 (85.5%) | Ref | | 302 (74.6%) | Ref | |
| 19-24 | 283 (97.3%) | 6.0 (2.6 - 15.1) | < 0.001 | 436 (89.7%) | 3.0 (2.1 - 4.3) | < 0.001 |
| **The highest education level** | | | | | | |
| Did not complete primary school | 111 (89.5%) | Ref | | 148 (81.8%) | Ref | |
| Completed primary school | 198 (94.3%) | 1.9 (0.8 - 4.4) | 0.11 | 375 (81.3%) | 1.0 (0.6 - 1.5) | 0.90 |
| Completed secondary school or higher | 74 (100%) | -- | -- | 215 (86.3%) | 1.4 (0.8 - 2.4) | 0.20 |
| **Currently receiving formal education** | | | | | | |
| No | 351 (93.6%) | Ref | | 570 (86.6%) | Ref | |
| Yes | 32 (97%) | 2.2 (0.4 - 39.7) | 0.45 | 168 (72.1%) | 0.4 (0.3 - 0.6) | < 0.001 |
| **Health-system engagement** | | | | | | |
| **Ever pregnant** | | | | | | |
| No | 157 (90.2%) | Ref | | 489 (77.4%) | Ref | |
| Yes | 226 (96.6%) | 3.1 (1.3 - 7.7) | 0.011 | 249 (96.1%) | 7.3 (4.0 - 15) | < 0.001 |
| **Treated STI last 1 year** | | | | | | |
| No | 293 (92.4%) | Ref | | 619 (81.6%) | Ref | |
| Yes | 90 (98.9%) | 7.4 (1.5 - 132.7) | 0.052 | 119 (90.2%) | 2.1 (1.2 - 3.9) | 0.018 |
| **Programme engagement** | | | | | | |
| Not aware of HIV services | 279 (92.1%) | Ref | | 611 (81.6%) | Ref | |
| Awareness of HIV services | 47 (100%) | -- | 0.99 | 71 (88.8%) | 1.8 (0.9 - 3.9) | 0.11 |
| Ever contacted by peers/staff from an NGO/CBO | 21 (100%) | -- | 0.99 | 32 (91.4%) | 2.4 (0.8 - 10.1) | 0.15 |
| Registered with NGO/CBO | 36 (97.3%) | 3.1 (0.6 - 56.2) | 0.28 | 24 (88.9%) | 1.8 (0.6 - 7.7) | 0.34 |
| **Sexual behavior and risk perception** | | | | | | |
| **Duration of sexual activity[c]** | | | | | | |
| <2 years | 54 (85.7%) | Ref | | 270 (72.8%) | Ref | |
| >=2 years | 329 (95.4%) | 3.4 (1.4 – 8.0) | 0.005 | 468 (90.0%) | 3.4 (2.3 – 4.9) | < 0.001 |
| **Duration in sex work** | | | | | | |
| <2 years | 185 (92.5%) | Ref | | -- | -- | -- |
| >=2 years | 198 (95.2%) | 1.6 (0.7 - 3.8) | 0.26 | -- | -- | -- |
| **Self-assessed risk of HIV acquisition[d]** | | | | | | |
| No risk at all/small/unsure | 167 (92.3%) | Ref | | 463 (82.1%) | Ref | |
| Moderate/Great | 211 (95%) | 1.6 (0.7 - 3.7) | 0.30 | 264 (83.5%) | 1.1 (0.8 - 1.6) | 0.59 |

Abbreviations: CBO (community-based organization); CI (confidence interval); NGO (non-governmental organization); NSW (young women not engaged in sex work); OR (odds ratio); STI (sexually transmitted infection); YSW (young women who sell sex)

[a]Physical establishments hotspots include bars, night clubs, hotels, guest houses, lodges, restaurants, local brew dens, sex dens and brothels
[b]Public spaces hotspots include streets and other public places
[c]N=55/1299 missing was imputed by adjusting for age at the interview
[d]Excluding individuals who disclosed they are living with HIV

**Table 3A. Univariate and multivariable analyses of factors associated with the history of HIV testing among adolescent girls and young women aged 14-24 years in Mombasa, Kenya (N=1299).**

| | Ever received HIV test | | | |
|---|---|---|---|---|
| | Crude OR (95% CI) | p-value | Adjusted OR (95% CI)[e] | p-value |
| **Socio-demographic characteristics** | | | | |
| **Engagement in sex work** | | | | |
| No | Ref | | Ref | |
| Yes | 3.2 (2.1 – 5.0) | < 0.001 | 1.5 (0.9 – 2.5) | 0.09 |
| **Type of recruitment hotspot** | | | | |
| Physical establishments[a] | 0.8 (0.5 - 1.2) | 0.24 | -- | |
| Public spaces[b] | Ref | | -- | |
| **Age in years** | | | | |
| 14-18 | Ref | < 0.001 | Ref | 0.002 |
| 19-24 | 3.7 (2.7 – 5.2) | | 1.8 (1.2 - 2.6) | |
| **Completed primary school** | | | | |
| No | Ref | 0.42 | -- | |
| Yes | 1.2 (0.8 - 1.7) | | -- | |
| **Currently receiving formal education** | | | | |
| No | Ref | < 0.001 | Ref | 0.30 |
| Yes | 0.4 (0.3 - 0.5) | | 0.8 (0.6 - 1.2) | |
| **Health-system engagement** | | | | |
| **Ever pregnant** | | | | |
| No | Ref | < 0.001 | Ref | < 0.001 |
| Yes | 6.4 (4.1 – 11.2) | | 3.6 (2.2 - 6.3) | |
| **Treated STI last 1 year** | | | | |
| No | Ref | < 0.001 | Ref | 0.047 |
| Yes | 2.7 (1.6 – 4.9) | | 1.8 (1.0 - 3.4) | |
| **Awareness of HIV services** | | | | |
| No | Ref | < 0.001 | Ref | 0.002 |
| Yes | 2.6 (1.6 – 4.6) | | 2.4 (1.4 - 4.3) | |
| **Sexual behavior and risk perception** | | | | |
| **Duration of sexual activity**[c] | | | | |
| <2 years | Ref | < 0.001 | Ref | < 0.001 |
| >=2 years | 4.0 (2.9 - 5.5) | | 2.1 (1.4 – 3.0) | |
| **Duration in sex work**[f] | | | | |
| <2 years | -- | | -- | |
| >=2 years | -- | | -- | |
| **Self-assessed risk of HIV acquisition**[d] | | | | |
| No risk at all/small/unsure | Ref | 0.057 | Ref | 0.90 |
| Moderate/Great | 1.4 (1.0 - 1.9) | | 1.0 (0.7 – 1.5) | |

Abbreviations: CI (confidence interval); NSW (young women not engaged in sex work); OR (odds ratio); STI (sexually transmitted infection); YSW (young women who sell sex)
[a]Physical establishments hotspots include bars, night clubs, hotels, guest houses, lodges, restaurants, local brew dens, sex dens and brothels
[b]Public spaces hotspots include streets and other public places
[c]N=55/1299 missing was imputed by adjusting for age at the interview
[d]Excluding individuals who disclosed they are living with HIV
[e]Not included covariates in multivariable analysis if significance level > 0.1 in univariate analysis
[f]Not included in univariable and multivariable analysis due to duration in sex work only applied to YSW

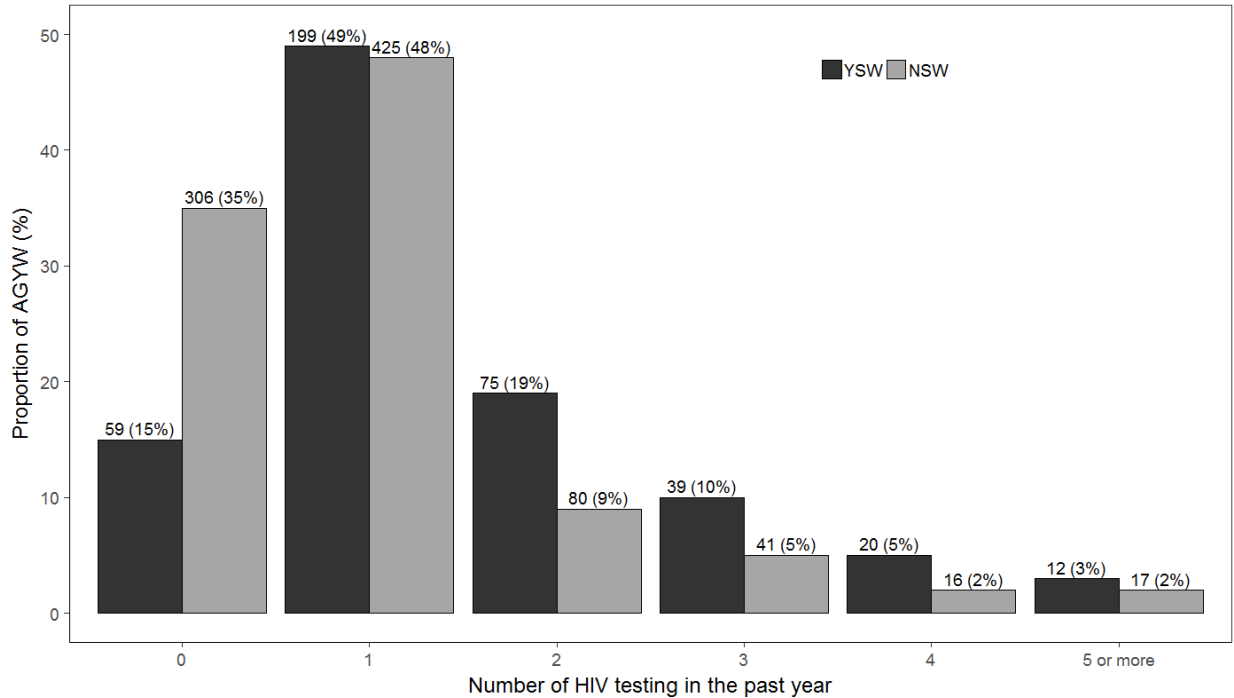

**Figure 1A. Frequency of HIV testing in the past year among adolescent girls and young women (AGYW) aged 14-24 years engagement in sex work in Mombasa, Kenya (N=1289).**
Abbreviations: YSW (young women who sell sex); NSW (young women not engaged in sex work).
Population excludes individuals who were diagnosed with HIV >1 year ago.

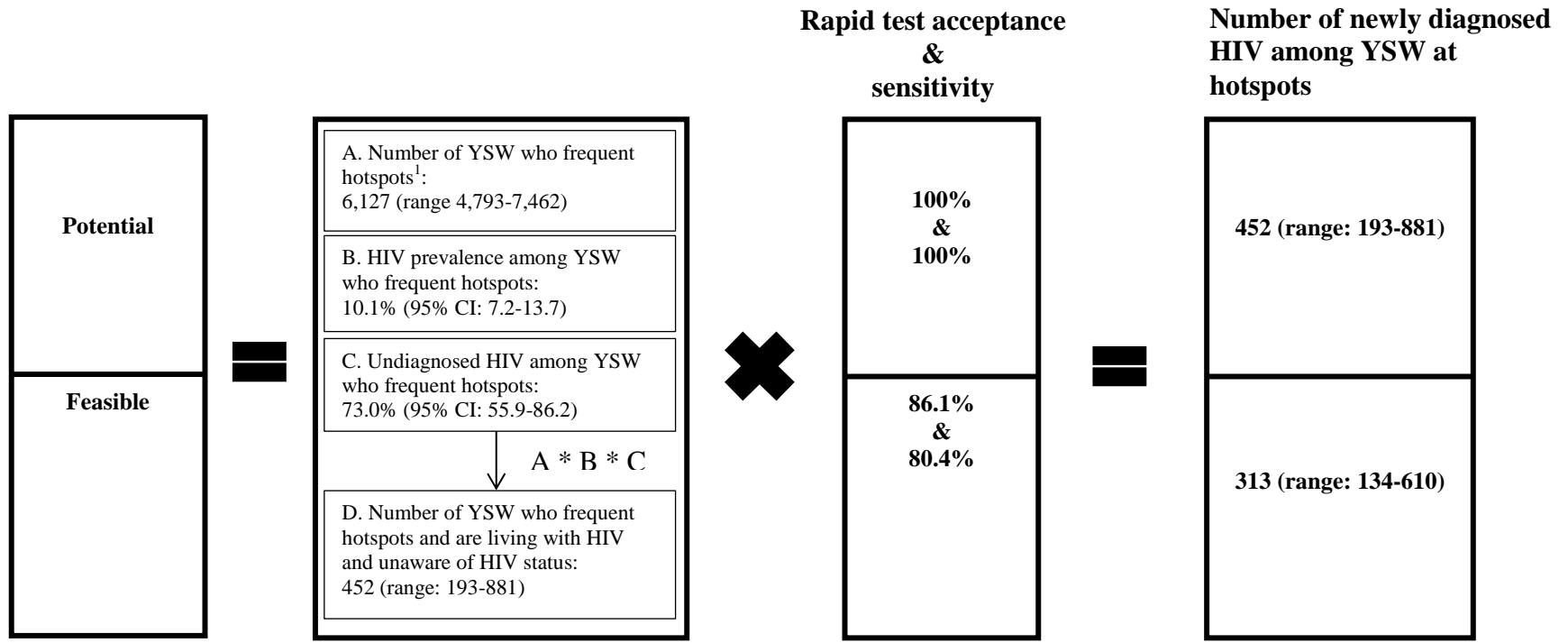

**Figure 2A. Triangulating the number of young women who sell sex living with HIV who could be diagnosed via hotspot-based HIV testing strategy in Mombasa, Kenya.**
Abbreviation: YSW (young women who sell sex).
[1]Cheuk E, Isac S, Musyoki H, et al. Informing HIV Prevention Programs for Adolescent Girls and Young Women: A Modified Approach to Programmatic Mapping and Key Population Size Estimation. JMIR Public Health Surveill. 2019;5(2):e11196.

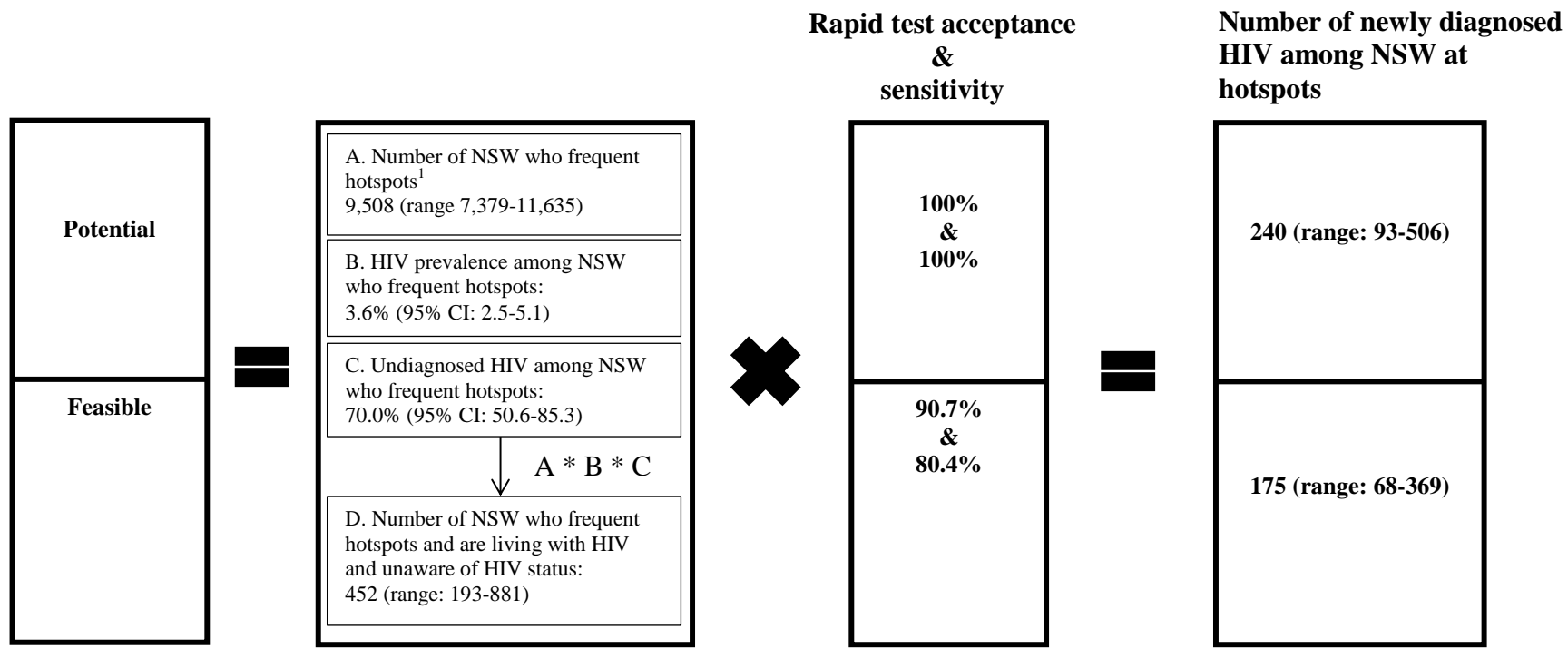

**Figure 3A. Triangulating the number of young women who sell sex living with HIV who could be diagnosed via hotspot-based HIV testing strategy in Mombasa, Kenya.**
Abbreviation: NSW (young women not engaged in sex work).
[1]Cheuk E, Isac S, Musyoki H, et al. Informing HIV Prevention Programs for Adolescent Girls and Young Women: A Modified Approach to Programmatic Mapping and Key Population Size Estimation. JMIR Public Health Surveill. 2019;5(2):e11196.